\documentstyle[11pt,a4,graphicx]{article}

\makeatletter

\ifx\normalfont\undefined
\def\normalfont{\reset@font}
\fi
\ifx\sffamily\undefined
\def\sffamily{\sf}
\fi
\ifx\bfseries\undefined
\def\bfseries{\bf}
\fi

\def\section{\@startuppersection{section}{1}{\z@ }%
{-3.5ex plus-1ex minus -.2ex}%
{2.3ex plus.2ex}{\normalfont\normalsize\sffamily\bfseries}}

\def\subsection{\@startsection{subsection}{2}{\z@ }{-3.25ex plus-1ex %
minus-.2ex}{1.5ex plus.2ex}{\normalfont\normalsize\sffamily\bfseries}}

\def\@startuppersection#1#2#3#4#5#6{\if@noskipsec \leavevmode \fi
   \par \@tempskipa #4\relax
   \@afterindenttrue
   \ifdim \@tempskipa <\z@ \@tempskipa -\@tempskipa \@afterindentfalse\fi
   \if@nobreak \everypar{}\else
     \addpenalty{\@secpenalty}\addvspace{\@tempskipa}\fi \@ifstar
     {\@upperssect{#3}{#4}{#5}{#6}}{\@dblarg{\@uppersect{#1}{#2}{#3}{#4}{#5}{#6}}}}

\def\@upperssect#1#2#3#4#5{\@tempskipa #3\relax
   \ifdim \@tempskipa>\z@
     \begingroup #4\@hangfrom{\hskip #1}{\interlinepenalty \@M %
{\uppercase{#5}}\par}\endgroup
   \else \def\@svsechd{#4\hskip #1\relax #5}\fi
    \@xsect{#3}}

\def\@uppersect#1#2#3#4#5#6[#7]#8{\ifnum #2>\c@secnumdepth
     \let\@svsec\@empty\else
     \refstepcounter{#1}\edef\@svsec{\csname the#1\endcsname . }\fi
     \@tempskipa #5\relax
      \ifdim \@tempskipa>\z@
        \begingroup #6\relax
          \@hangfrom{\hskip #3\relax\@svsec}{\interlinepenalty \@M  %
	{\uppercase{#8}}\par}%
        \endgroup
       \csname #1mark\endcsname{#7}\addcontentsline
         {toc}{#1}{\ifnum #2>\c@secnumdepth \else
                      \protect\numberline{\csname the#1\endcsname}\fi
                    #7}\else
        \def\@svsechd{#6\hskip #3\relax  
                   \@svsec #8\csname #1mark\endcsname
                      {#7}\addcontentsline
                           {toc}{#1}{\ifnum #2>\c@secnumdepth \else
                             \protect\numberline{\csname the#1\endcsname}\fi
                       #7}}\fi
     \@xsect{#5}}

\def\@sect#1#2#3#4#5#6[#7]#8{\ifnum #2>\c@secnumdepth
     \let\@svsec\@empty\else
     \refstepcounter{#1}\edef\@svsec{\csname the#1\endcsname . }\fi
     \@tempskipa #5\relax
      \ifdim \@tempskipa>\z@
        \begingroup #6\relax
          \@hangfrom{\hskip #3\relax\@svsec}{\interlinepenalty \@M  %
	#8\par}%
        \endgroup
       \csname #1mark\endcsname{#7}\addcontentsline
         {toc}{#1}{\ifnum #2>\c@secnumdepth \else
                      \protect\numberline{\csname the#1\endcsname}\fi
                    #7}\else
        \def\@svsechd{#6\hskip #3\relax  
                   \@svsec #8\csname #1mark\endcsname
                      {#7}\addcontentsline
                           {toc}{#1}{\ifnum #2>\c@secnumdepth \else
                             \protect\numberline{\csname the#1\endcsname}\fi
                       #7}}\fi
     \@xsect{#5}}

\def\theequation{\arabic{section}.\arabic{equation}}

\def\bio#1#2#3#4#5{{\rm #1, #2 }{\bf #3}{\rm, #4 (#5)}}

\def\thebibliography#1{

\pagebreak
\section*{References}
\list
 {\leavevmode\raise1ex\hbox{\tiny\arabic{enumi}}}%
 {\settowidth\labelwidth{\tiny #1}\leftmargin\labelwidth
 \itemsep=0pt\labelsep=0pt
 \usecounter{enumi}}
 \sloppy\clubpenalty4000\widowpenalty4000
 \sfcode`\.=1000
 \normalfont\small
\relax}

\def\@cite#1#2{{#1\if@tempswa , #2\fi}}

\def\citenote#1{\raise 1ex\hbox{\tiny{\cite{#1}}}}

\def\abstract{\if@twocolumn
\section*{Abstract}
\else \small 
\quotation\noindent 
\fi}
\def\endabstract{\if@twocolumn\else\endquotation\fi}

\def\@maketitle{
 \null
 \vskip 2em 
 {\normalfont\LARGE\sffamily\bfseries\noindent \@title \par} 
\vskip 1.5em {\sf \lineskip .5em
\quotation\noindent\@author\endquotation\par} 
 \vskip 1em 
 \par
 \vskip 1.5em} 

\newcounter{equation@backup}
\def\mathletters{%
\addtocounter{equation}{1}
\edef\@currentlabel{\theequation}
\setcounter{equation@backup}{\value{equation}}
\setcounter{equation}{0}
\def\theequation{\arabic{section}.\arabic{equation@backup}\alph{equation}}
}

\def\preprintlabel#1{\gdef\@preprintlabel{#1}}
\def\preprintdate#1{\gdef\@preprintdate{#1}}

\let\@oldendabstract=\endabstract
\def\endabstract{%
\@oldendabstract
\vfill
\noindent\@preprintlabel\hfill\penalty-10000
\noindent\@preprintdate\hfill\penalty-10000
\pagebreak
}

\def\maketitle{\thispagestyle{empty}
 \let\footnotesize\small
 \def\thefootnote{\fnsymbol{footnote}}
 \setcounter{page}{0}%
 \@maketitle
\@thanks
\setcounter{footnote}{0}%
\let\thanks\relax
\gdef\@thanks{}\gdef\@author{}\gdef\@title{}\let\maketitle\relax}

\makeatother

\preprintlabel{LPTHE Orsay 96-98}
\preprintdate{Decembre 1996}

\def\titre{Two-component plasma in a gravitational field}

\ifx\mathbf\undefined 
\newcommand{\mathbf}[1]{\mbox{\bf #1}}
\fi
\ifx\mathrm\undefined 
\newcommand{\mathrm}[1]{\mbox{\rm #1}}
\fi

\newcommand{\mbf}[1]{{\mathbf #1}}
\def\r{\mbf{r}}

\newcommand{\one}{\mathbf{1}\kern-.46ex\rule{.07ex}{1.45ex}%
\rule{.3ex}{.1ex}\kern.3ex{}}

\newcommand{\dslash}{\hskip.1ex\hbox to 0pt{/\hss}\kern-.1ex\partial}

\newcommand{\moko}{\frac{m_0}{k_0}}
\newcommand{\dk}{\delta\!k}
\newcommand{\Ass}{A_{\sigma'}^{\sigma}}
\newcommand{\Bss}{B_{\sigma'}^{\sigma}}

\begin{document}
\title{\titre}
\author{Gabriel T\'ellez\thanks{e-mail address: tellez@stat.th.u-psud.fr}\\%
{\rm \small \it Laboratoire de Physique Th\'eorique et Hautes Energies}%
\thanks{Laboratoire associ\'e au Centre National de la Recherche
Scientifique - URA D0063} \\
{\rm\small\it Universit\'e de Paris-Sud,
91405 Orsay, France}
}
\date{}
\maketitle
\begin{abstract}
In this paper we study a model for the sedimentation equilibrium
of a charged colloidal suspension: the
two-dimensional two-component plasma in a gravitational field which is
exactly solvable at a special value of the reduced inverse temperature
$\Gamma=2$. The density profiles are
computed. 
The heavy particles accumulate at the bottom of the cointainer. If the
container is high enough, an excess of light counterions form a cloud
floating at some altitude.
\end{abstract}
\setcounter{equation}{0}
\section{Introduction}

When colloidal particles are solvated in a polar fluid they usually
release counterions and therefore they acquire an electric charge. Then,
a solution of heavy charges (the colloidal particles) and light charges
(the counterions) is obtained. A model of such charged colloidal
suspensions in sedimentation equilibrium based on the local density
approximation has been proposed\citenote{BibHans} and numerical
computations about that model show that the heavy charges accumulate at
the bottom of the container while the light particles accumulate at the
top.  The suspension behaves like a condenser: a vertical electric field
is induced into the suspension. Also, several experimental measurements
of the densities profiles in such suspensions have been done.

In this paper we consider a simple model of sedimentation equilibrium,
taking into account only the Coulomb interactions between the particles
and the gravitational force (in particular we neglect excluded volume
interactions), which is exactly solvable. The model is a two-dimensional
two-component plasma in a gravitational field, that is a system composed
of two species of particles with electric charges $\pm q$ and masses
$M_{\pm}$. The Coulomb interaction is logarithmic in two-dimensions, and
the particles are submitted to a uniform vertical gravitational field.

The two-component plasma model has been solved by Gaudin\citenote{Gaudin}
when the inverse temperature $\beta$ verifies $\Gamma:=\beta q^2=2$, and
much work has been done with that model. Studing the electrical
double layer for the two-component plasma, Cornu and
Jancovici\citenote{CorJan} introduced a very general method for treating the
two-component plasma in presence of an external potential. We use here
their method to deal with the gravitational field.

In the following section we present the model and recall some results of
the method as given in Ref.~\cite{CorJan}. In section~\ref{Results}, we solve
the model and 
compute quantities such as the density profiles and the electric field in
the suspension.

\section{The model and the general method of resolution}
\label{Model}
\setcounter{equation}{0}

The model is a two-dimensional system of particles of charges $\pm q$
and masses $M_\pm$. The position $\mbf{r}$ of a particle is represented
by its Cartesian coordinates $(x,y)$.  The gravitational field is
$\mbf{g}=-g\hat{\mbf{y}}$. The particles are in a container of height
$h$ and infinite width. The bottom of the container is at $y=0$.  The
half-spaces $y<0$ and $y>h$ are impenetrable to the particles.  The
interaction of the particles with the gravitational field is
\begin{equation}
\label{mgz}
M_+ g \sum_i y_i^+ + M_- g \sum_j y_j^-
\end{equation}
where $y_i^\pm$ is the altitude of the $i^{\mbox{\scriptsize th}}$ particle of
charge $\pm q$. Expression~(\ref{mgz}) can be rewritten as
\begin{equation}
M_{\mbox{\scriptsize eff}}\, g \left( \sum_i y_i^+ +\sum_j y_j^- \right)
- qE_{\mbox{\scriptsize eff}} \left( \sum_i y_i^+ - \sum_j y_j^- \right)
\end{equation}
where $M_{\mbox{\scriptsize eff}}=(M_+ + M_-)/2$ and $qE_{\mbox{\scriptsize
eff}}=-(M_+-M_-)g/2$. This shows that the system is equivalent to a
system where all the particles have the same mass $M_{\mbox{\scriptsize eff}}$
but there is an external electric field $E_{\mbox{\scriptsize eff}}$ acting on
the particles. For this reason it is useful to define $k_\pm=\beta
M_\pm g$, $k_0=(k_++k_-)/2$ and $\dk=k_+-k_-$. 

For $\Gamma
\geq 2$ a point-particle model is unstable against the collapse of pairs of
particles of opposite charge. To prevent this collapse we introduce some
short range cutoff by representing the particles as hard discs of
diameter $R$ and obtain results in the small-$R$ limit. In fact, at
$\Gamma=2$, only the one-body density diverges, while the $n$-body
correlations ($n\geq2$) have well-defined limits as $R\to0$.  From now
on we shall consider that the temperature is such that $\Gamma=2$.  The
model is solved in the grand canonical ensemble. It is convenient to
introduce position-dependent rescaled fugacities that have inverse
length dimensions: $m_\pm (\mbf{r})=m_0 \exp(-k_\pm y)$. There are two
length scales in this problem: the gravitational length given by
$k_0^{-1}$ and the screening length which in fact is
given\citenote{CorJan} by the inverse of the rescaled fugacity
$m_0^{-1}$. Here we concentrate on the usual case close to the
continuous medium limit where $m_0\gg k_0$.

Using the equivalence between the
two-dimensional Coulomb gas at $\Gamma=2$ and the free Dirac
field\citenote{CorJan,Coleman,Samuel}, one can write
 the grand
partition function in the form
\begin{equation}
\Xi=\det\left[\left(\dslash + m_+(\mbf{r})\frac{1+\sigma_z}{2}+
m_-(\mbf{r})\frac{1-\sigma_z}{2}\right){\dslash}^{-1}\right]\,,
\end{equation}
where $\dslash=\sigma_x\partial_x+\sigma_y\partial_y$ is the Dirac
operator in two dimensions and $\sigma_x$, $\sigma_y$ and $\sigma_z$ are
the Pauli matrices. The particle densities $\rho_s$ (where $s=\pm1$) 
and the truncated
two-body densities $\rho_{ss'}^{(2)T}$ can be obtained from
the Green functions $G_{ss'}$ defined by
\begin{equation}
\left[
 \dslash + m_+(\mbf{r})\frac{1+\sigma_z}{2}+
m_-(\mbf{r})\frac{1-\sigma_z}{2}
\right]G=\one\delta(\r-\r')\,.
\end{equation}
One finds
\begin{equation}
\label{defrho}
\rho_s(\mbf{r})=m_s(\mbf{r})G_{ss}(\mbf{r},\mbf{r})\,,
\end{equation}
and
\begin{equation}
\rho_{ss'}^{(2)T}(\r,\r')=-m_s(\r)m_{s'}(\r')G_{ss'}(\r,\r')G_{s's}(\r',\r)\,.
\end{equation}

It is convenient to write the fugacities as 
$m_\pm(\mbf{r})=m(\mbf{r})\exp[-(\pm)2V(\mbf{r})]$ where
$m(\mbf{r})=m_0\exp[-k_0 y]$ and $V(\mbf{r})=\dk\,y/4$, and
to define
\begin{equation}
g_{ss'}(\mbf{r},\mbf{r}')=\exp[-sV(\mbf{r})]G_{ss'}(\mbf{r},\mbf{r}')\exp[-s'V(\mbf{r}')]\,.
\end{equation}
Then, in terms of the operators
$a=\partial_x+i\partial_y+\partial_xV+i\partial_yV$ and
$a^{\dag}=-\partial_x+i\partial_y+\partial_xV-i\partial_yV$, the
functions $g_{ss'}$ are given by the partial differential equations
\begin{equation}
\label{Gengpp}
\left[m(\mbf{r})+a^{\dag}
\left(m(\mbf{r})\right)^{-1}a\right]
g_{++}(\mbf{r},\mbf{r}')=\delta(\mbf{r}-\mbf{r}')\,,
\end{equation}
\begin{equation}
\label{Gengmm}
\left[m(\mbf{r})+a\left(m(\mbf{r})\right)^{-1}
a^{\dag}\right]g_{--}(\mbf{r},\mbf{r}')=
\delta(\mbf{r}-\mbf{r}')\,,
\end{equation}
and
\begin{equation}
\label{Gengmp}
g_{-+}(\mbf{r},\mbf{r}')=
-\left[m(\mbf{r})\right]^{-1}ag_{++}(\mbf{r},\mbf{r}')\,,
\end{equation}
\begin{equation}
\label{Gengpm}
g_{+-}(\mbf{r},\mbf{r}')=
\left[m(\mbf{r})\right]^{-1}a^{\dag}g_{--}(\mbf{r},\mbf{r}')\,.
\end{equation}

\setcounter{equation}{0}
\section{Calculations and results}
\label{Results}
\subsection{The Green functions}
Because of the translational invariance along the $x$ axis we can
suppose that $x'=0$. It is useful
to work with the Fourier transform $\hat{g}_{ss'}$ 
of Green functions $g_{ss'}$ defined by
\begin{equation}
\label{Fourierg}
g_{ss'}(\r,\r')=\frac{1}{2\pi}\int_{-\infty}^{+\infty}\hat{g}_{ss'}(y,y',k)
e^{ikx}\,dk\,.
\end{equation}
From equation~(\ref{Gengpp}) we find that the Fourier transform
$\hat{g}_{++}$ satisfies the ordinary differential equation
\begin{equation}
\label{gppchapeau}
\begin{array}{ccc}
\lefteqn{
\left[(k+\dk/4)k_0-(k+\dk/4)^2-m_0^2e^{-2k_0y}\right]
\hat{g}_{++}(y,y',k)}\qquad\qquad\qquad
\\
 & & +k_0\partial_y\hat{g}_{++}(y,y',k)+\partial_y^2 \hat{g}_{++}(y,y',k)
=-m_0e^{-k_0y}\delta(y-y')\,.
\end{array}
\end{equation}
In terms of $u=\exp[-k_0y]$, $u'=\exp[-k_0y']$ 
and $f_{ss'}(u,u',k)=\hat{g}_{ss'}(y,y',k)$, we have for
$f_{++}$ the equation
\begin{equation}
\label{Eqfpp}
\partial^2_u f_{++}(u,u',k)
-\left[(\nu_+^2-1/4)u^{-2}+\left(\frac{m_0}{k_0}\right)^2
\right]f_{++}(u,u',k)=-\frac{m_0}{k_0}\delta(u-u')\,,
\end{equation}
where $\nu_+^2=[k-(k_-/2)]^2/k_0^2$.
A similar equation is obtained for $f_{--}$ by replacing $\nu_+^2$ by 
$\nu_-^2=[-k-(k_+/2)]^2/k_0^2$.
Equation~(\ref{Eqfpp}) can be solved in terms of the modified Bessel
functions $I_{\nu_+}$ and $K_{\nu_+}$:
\begin{equation}
\label{fpp}
f_{++}(u,u',k)=\frac{m_0}{k_0}\sqrt{uu'}
\left[
I_{\nu_+}(\moko u_<)K_{\nu_+}(\moko u_>)+ A I_{\nu_+}(\moko u)
+B K_{\nu_+}(\moko u)
\right]\,,
\end{equation}
where $u_<=\min(u,u')$, $u_>=\max(u,u')$, and $A$
and $B$ are constants (with respect to $u$) that depend on the boundary
conditions. Note that the sign of $\nu_+$ can be arbitrarly chosen; we
choose~$\nu_+$ positive.
Using equation~(\ref{Gengmp}) we find
\begin{eqnarray}
\lefteqn{f_{-+}(u,u',k)=-i\frac{m_0}{k_0}\sqrt{uu'}
\,\Bigg[ 
A I_{\nu_++\sigma}(\moko u)
-B K_{\nu_++\sigma}(\moko u)}
\label{fmp}
\hskip10em\\
&&
+\cases{
I_{\nu_++\sigma}(\moko u)K_{\nu_+}(\moko u'), &if $u<u'$\cr
-I_{\nu_+}(\moko u')K_{\nu_++\sigma}(\moko u), &if $u'<u$\cr
}
\Bigg]\,,\nonumber
\end{eqnarray}
where $\sigma$ is the sign of $k-(k_-/2)$. 

For a source point in the allowed region $0<y'<h$,
in the impenetrable region $y<0$ the particle fugacities vanish and the
equations for the Fourier transforms of $G_{++}$ and $G_{-+}$ reduce to
\begin{equation}
\begin{array}{cc}
(k+\partial_y)\hat{G}_{++}(y,y',k)=0\,,
&
(k-\partial_y)\hat{G}_{-+}(y,y',k)=0\,,
\end{array}
\end{equation}
with the solution
\begin{equation}
\begin{array}{cc}
\label{SolGmenos}
\hat{G}_{++}(y,y',k)=Ce^{-ky}\,,&\hat{G}_{-+}(y,y',k)=De^{ky}\,.
\end{array}
\end{equation}
Similar equations hold for $y>h$.

The boundary conditions are that $\hat{g}_{ss'}$ must vanish when
$y\to\pm\infty$ and be continuous at $y=0$ and at $y=h$.  The boundary
conditions when $y\to-\infty$ and equations~(\ref{SolGmenos}) lead to
$f_{++}(1,u',k)=0$ if $k>0$ and $f_{-+}(1,u',k)=0$ if $k<0$, while the
boundary conditions when $y\to+\infty$ give at $y=h$:
$f_{++}(u_h,u',k)=0$ if $k<0$ and $f_{-+}(u_h,u',k)=0$ if $k>0$, where
we have defined $u_h=\exp(-k_0h)$.  Using these boundary conditions
in~(\ref{fpp}) and~(\ref{fmp}) we find the values of $A$ and $B$ which
depend on the sign of $k$ and the sign of $k-k_-/2$. We shall use the
following notation: $\Ass(u',\nu_+)$ and $\Bss(u',\nu_+)$ are the values
of $A$ and $B$ when $k$ has sign $\sigma'$ and $k-k_-/2$ has sign
$\sigma$.  Using this notation we have
\begin{mathletters}
\begin{equation}  
A_+^{\sigma}(u',\nu)=%
-K_\nu(\moko)\frac{%
I_\nu(\moko u')K_{\nu+\sigma}(\moko u_h)+
K_\nu(\moko u')I_{\nu+\sigma}(\moko u_h)}{%
I_\nu(\moko)K_{\nu+\sigma}(\moko u_h)+
K_\nu(\moko)I_{\nu+\sigma}(\moko u_h)}\,,
\end{equation}
\begin{equation}
B_+^{\sigma}(u',\nu)=%
I_{\nu+\sigma}(\moko u_h)\frac{%
I_\nu(\moko)K_\nu(\moko u')-
K_\nu(\moko)I_\nu(\moko u')}{%
I_\nu(\moko)K_{\nu+\sigma}(\moko u_h)+
K_\nu(\moko)I_{\nu+\sigma}(\moko u_h)}\,,
\end{equation}
\begin{equation}
A_-^{\sigma}(u',\nu)=
K_{\nu+\sigma}(\moko)\frac{%
K_\nu(\moko u_h)I_\nu(\moko u')-
I_\nu(\moko u_h)K_\nu(\moko u')}{%
K_\nu(\moko u_h)I_{\nu+\sigma}(\moko)+
I_\nu(\moko u_h)K_{\nu+\sigma}(\moko)}\,,
\end{equation}
\begin{equation}
B_-^{\sigma}(u',\nu)=
-I_{\nu}(\moko u_h)\frac{%
K_{\nu+\sigma}(\moko)I_\nu(\moko u')+
I_{\nu+\sigma}(\moko)K_\nu(\moko u')}{%
K_\nu(\moko u_h)I_{\nu+\sigma}(\moko)+
I_\nu(\moko u_h)K_{\nu+\sigma}(\moko)}\,.
\end{equation}
\end{mathletters}
Similar calculations can be done for $f_{--}$. Then changing the integral
over $k$ in~(\ref{Fourierg}) into an integral over $\nu$ we find for the
Green functions
\begin{eqnarray}
\lefteqn{g_{ss}(\r,\r')=\frac{m_0}{2\pi}\exp[isk_{-s}x/2]\sqrt{uu'}}
\qquad\qquad\quad&\quad\ 
\nonumber\\
&\lefteqn{\times\Bigg[
}&2
\int_0^{+\infty} I_{\nu}(\moko u_<)K_{\nu}(\moko
u_>)\cos (k_0x\nu)
\,d\nu
\nonumber\\
\label{gssint}
&&+
\int_0^{+\infty} \left[A_+^+(u',\nu)I_\nu(\moko u) 
+B_+^+(u',\nu)K_\nu(\moko u)\right]
e^{isk_0x\nu}d\nu\\
&&+
\int_0^{k_{-s}/2k_0} \left[A_+^-(u',\nu)I_\nu(\moko u)
+B_+^-(u',\nu)K_\nu(\moko u)\right]
e^{-isk_0x\nu}d\nu
\nonumber\\
&&+
\int_{k_{-s}/2k_0}^{+\infty}\left[ 
A_-^-(u',\nu) I_\nu(\moko u) +
B_-^-(u',\nu)K_\nu(\moko u)\right]
e^{-isk_0x\nu}d\nu
\,\,\Bigg]\,.\nonumber
\end{eqnarray}
while $g_{-+}$ and $g_{+-}$ are given by (\ref{Gengmp}) and (\ref{Gengpm}).

An interesting limit is when the height $h$ of the container goes to
infinity. One can take the limits of $\Ass(u',\nu)$ and~$\Bss(u',\nu)$
when $u_h\to0$ in~(\ref{gssint}). One finds
\begin{eqnarray}
\lefteqn{g_{ss}(\r,\r')=\frac{m_0}{2\pi}\exp[isk_{-s}x/2]\sqrt{uu'}}
\qquad\qquad\quad&\quad
\nonumber\\
&\lefteqn{\times\Bigg[
}&2
\int_0^{+\infty} I_{\nu}(\moko u_<)K_{\nu}(\moko
u_>)\cos(k_0x\nu)
\,d\nu
\nonumber\\
&&-
\int_0^{+\infty} 
I_\nu(\moko u) I_\nu(\moko u') 
\frac{K_\nu(\moko)}{I_\nu(\moko)}
e^{isk_0x\nu}d\nu
\nonumber\\
\label{gss}
&&+
\int_{k_{-s}/2k_0}^{+\infty}
I_\nu(\moko u) I_\nu(\moko u') 
\frac{K_{\nu-1}(\moko)}{I_{\nu-1}(\moko)}
e^{-isk_0x\nu}d\nu
\\
&&+
\int_0^{k_{-s}/2k_0}\!\!\Bigg[
\!-\!I_\nu(\moko u)I_\nu(\moko u')
K_\nu(\moko)
\nonumber\\
&&\qquad\qquad\quad
+\Bigg\{
\!\!-\!\Big[I_\nu(\moko u)K_\nu(\moko u')
+K_\nu(\moko u)I_\nu(\moko u')\Big]
K_\nu(\moko) 
\nonumber\\
&&\qquad\qquad\qquad\quad\! 
+K_\nu(\moko u)K_\nu(\moko u')I_\nu(\moko) 
\Bigg\}\frac{2\sin\pi\nu}{\pi}
\Bigg]
\frac{e^{-isk_0x\nu}}{I_{-\nu}(\moko)}
d\nu
\Bigg].\nonumber
\end{eqnarray}

\subsection{The charge density}

\begin{figure}
\begin{center}
\includegraphics[width=\textwidth]{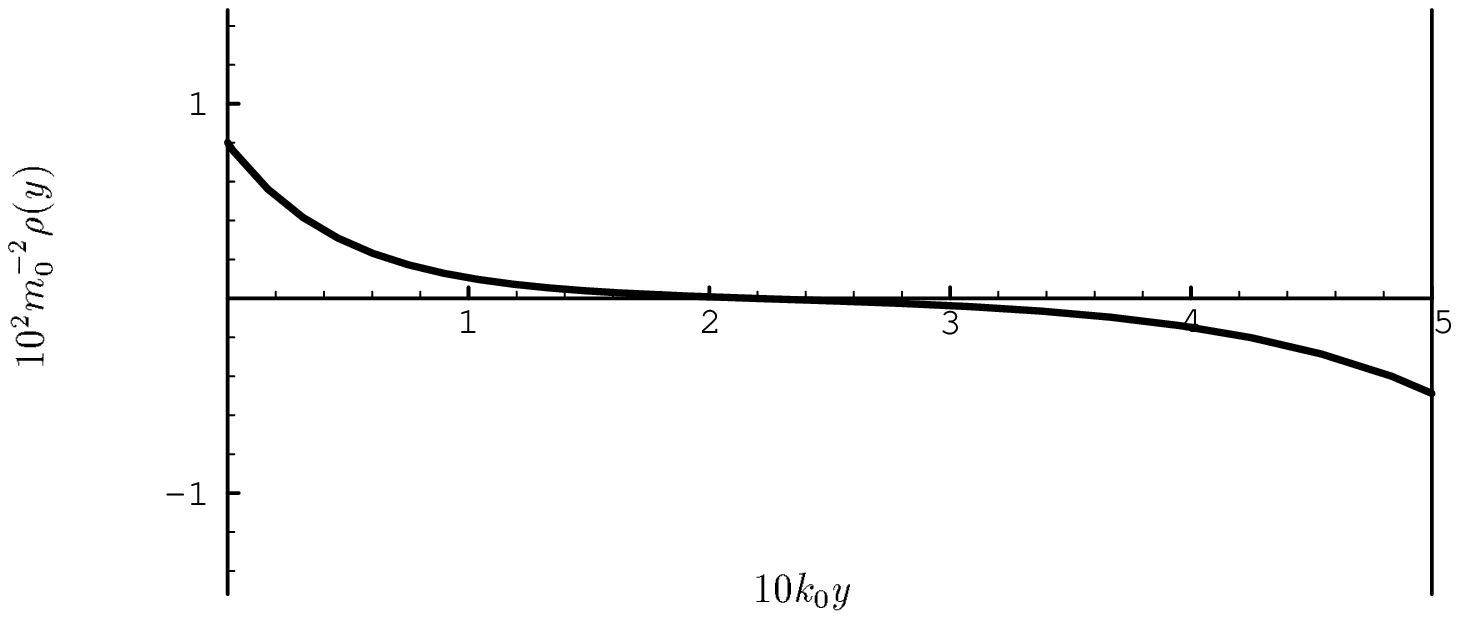}
\end{center}
\caption{\label{un}The charge density $\rho(y)$
for $m_0/k_0=10$,
$k_+=1.5k_0$, $k_-=0.5k_0$ and $h=0.5k_0^{-1}$.}
\end{figure}

\begin{figure}
\begin{center}
\includegraphics[width=\textwidth]{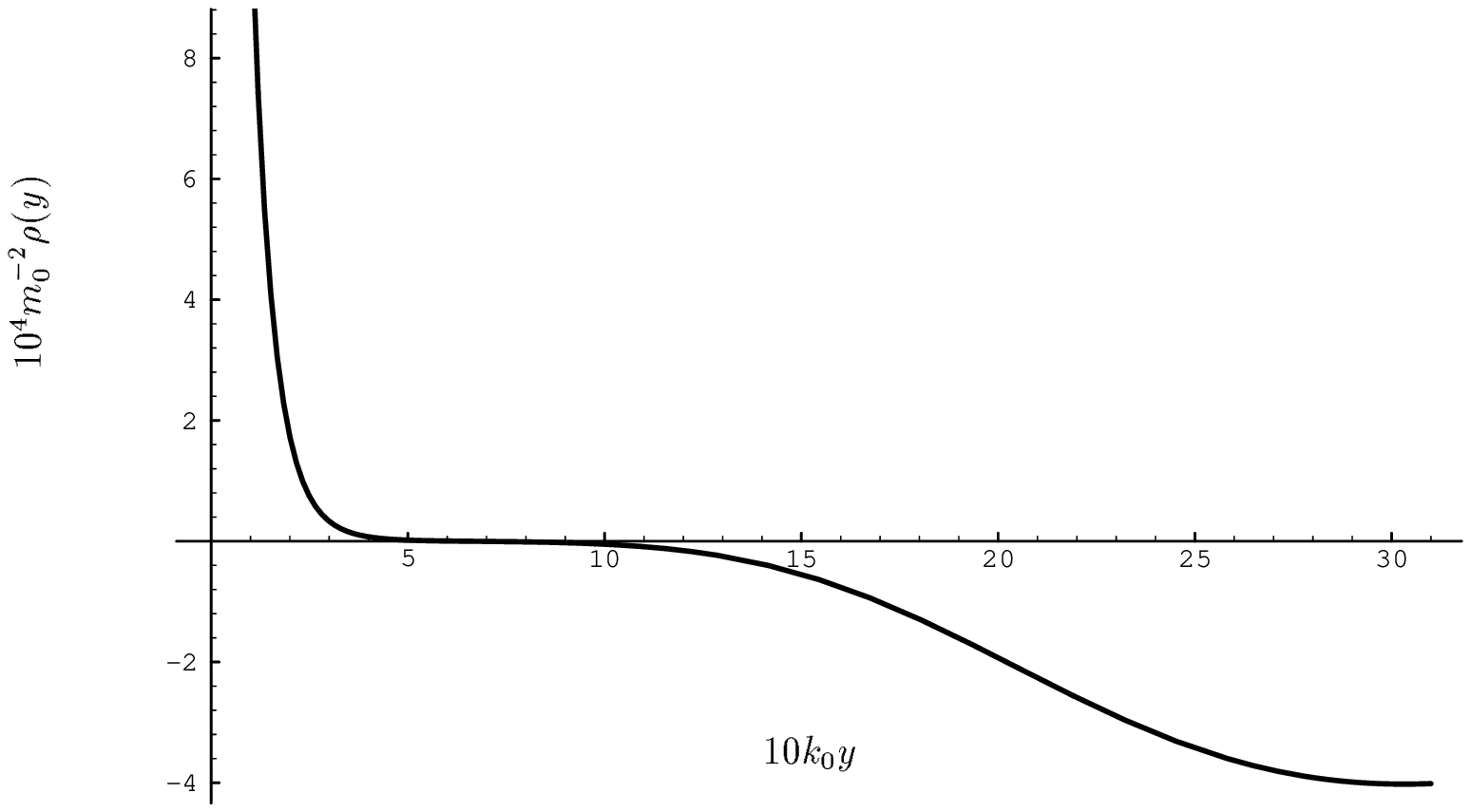}
\end{center}
\caption{\label{deux}
The charge density $\rho(y)$
for $m_0/k_0=10$,
$k_+=1.5k_0$, $k_-=0.5k_0$ and $h=3k_0^{-1}$.}
\end{figure}

\begin{figure}
\begin{center}
\includegraphics[width=\textwidth]{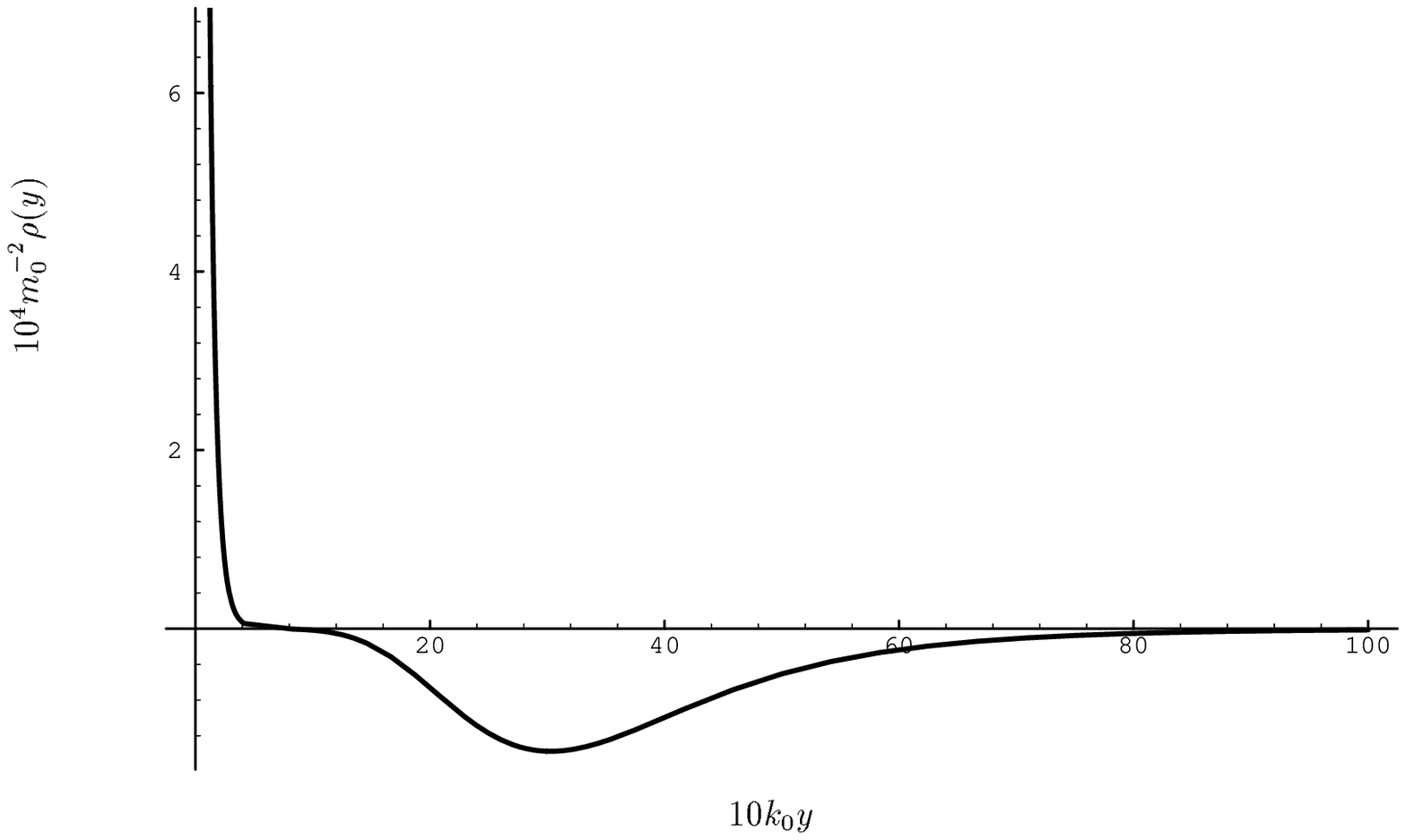}
\end{center}
\caption{The charge density $\rho(y)$
for $m_0/k_0=10$,
$k_+=1.5k_0$, $k_-=0.5k_0$ and $h=10k_0^{-1}$.
\label{trois}}
\end{figure}

The densities $\rho_{s}$ are given by equation~(\ref{defrho}). The first
integral in~(\ref{gssint}) and in~(\ref{gss}) diverges for $u'=u$ and
must be cut-off as mentioned in section~\ref{Model}; an equivalent
method is to introduce a cutoff $\nu_{\mbox{\scriptsize max}}$ in the
integral.  The other terms are finite when $u<1$ ($y\neq 0$).  The
charge density is $\rho=q(\rho_+-\rho_-)$ and has a finite limit when
the cut-off $R$ goes to zero.  Figures~\ref{un}, \ref{deux}
and~\ref{trois} show the charge density profile when the heavy
particles have charge $+q$, and for~$m_0/k_0=10$, $k_+=1.5k_0$,
$k_-=0.5k_0$ and different heights $h$ of the container ($k_0h=0.5$,
$3$, and~$10$ respectively).

When $h$ is of the same order or less than the gravitational length
$k_0^{-1}$ we notice a strong accumulation of heavy particles at the
bottom of the container and of light particles at the top, while in the
middle region the charge density is almost zero. In that zone there is an
almost constant electric field. The suspension behaves like a condenser.
This was noticed before in other models\citenote{BibHans}. Note that
we have considered the usual physical case where $m_0\gg k_0$. In the
hypothetical case where $k_0\gg m_0$ the neutral zone would not exist.
We shall assume from now on that $m_0\gg k_0$.

When $h\gg k_0^{-1}$, there still is a strong accumulation of heavy
particles at the bottom of the container but the light particles no
longer accumulate on the top of the container. There is an excess of
light particles at some altitude that will be shown later to be of order
$k_0^{-1}\ln(m_0/k_0)$.

Let us consider the case $h=+\infty$. The
charge density can be written as
\begin{eqnarray}
\rho(\r)=q\frac{m_0^2u^2}{2\pi}
\int_{k_{-}/2k_0}^{k_+/2k_0}&&
\hskip-1.6em\Bigg[ 
I_\nu(\moko u)^2
\frac{K_{\nu-1}(\moko)}{I_{\nu-1}(\moko)}
\nonumber\\
&&\hskip-.6em{}
+\Bigg\{   
I_\nu(\moko u)^2
K_\nu(\moko)
\\
&&\hskip1.6em{}
+\!\Big[\,
2\,I_\nu(\moko u)K_\nu(\moko u)
K_\nu(\moko) 
\nonumber\\
&&\hskip3.6em{}
-K_\nu(\moko u)^2I_\nu(\moko) 
\Big]\frac{2\sin\pi\nu}{\pi} 
\Bigg\} 
\frac{1}{I_{-\nu}(\moko)}\Bigg]
d\nu\,.\nonumber
\end{eqnarray}
First we shall study the charge density near the bottom of the container
and at intermediate altitudes. 
Remember that we are in the case $m_0\gg k_0$. If $y\ll
k_0^{-1}\ln(m_0/k_0)$ and $u>1/2$ then using the asymptotic expansion of
the Bessel functions we have
\begin{equation}
\label{rho0}
\rho(\r)\sim
-q\frac{m_0\dk}{4\pi}
\exp\left[-k_0y-2\moko (1-e^{-k_0 y})\right]\,.
\end{equation}
From this expression we notice that for $y\ll k_0^{-1}\ln(m_0/k_0)$,
the function $q\dk\,\rho(y)$ is a decreasing function of $y$.
Furthermore from expression~(\ref{rho0}) we see that the layer of heavy
particles has a thickness of order $m_0^{-1}$, the screening length.
Now at high altitudes for $y\gg k_0^{-1}\ln(m_0/k_0)$ we have
\begin{equation}
\rho(y)\sim-q\frac{m_0^2u^2}{\pi^2}
\int_0^{k_+/2k_0}-\int_0^{k_-/2k_0}
\sin(\pi\nu)K_\nu(\moko u)^2\,
d\nu\,.
\end{equation}
The asymptotic behavior when $y\to\infty$ of the integral
$\int_0^{k_s/2k_0}\sin(\pi\nu)K_\nu(\moko u)^2\,d\nu$ 
is computed in the Appendix. One
finds if $k_+\neq0$ and $k_-\neq0$
\begin{equation}
\label{rhoalinfini}
\rho(y)\sim -q\mathop{\mathrm{sgn}}(\dk) 
\frac{2^{k_>/k_0}}{8\pi^2}
\left[\moko\right]^{k_</k_0}
k_0\Gamma(k_>/2k_0)^2\sin(\pi k_>/2k_0)\,
\frac{e^{-k_< y}}{y}\,,
\end{equation}
where $k_<=\min(k_+,k_-)$ and $k_>=\max(k_+,k_-)$.
From expression~(\ref{rhoalinfini}), we see
that $q\dk\,\rho(y)$ is now an increasing function of $y$. This shows
that at an altitude of order $k_0^{-1}\ln(m_0/k_0)$ we a local minimum
of the function $q\dk\,\rho(y)$

The electric
field in the region where the system is almost neutral ($m_0^{-1}\ll y\ll
k_0^{-1}\ln(m_0/k_0)$)
can be estimated as follows. Remember that the system
is equivalent to a system of particles with the same mass in an external
electric field $E_{\mbox{\scriptsize eff}}=-(M_+-M_-)g/2q$. The
screening properties of this equivalent system will ensure that its
total electric field is zero, so the electric field in our system,
created by the accumulation of particles at the bottom and top of the
container (or at an altitude of order $k_0^{-1}\ln(m_0/k_0)$ if $h\gg
k_0^{-1}\ln(m_0/k_0)$), will be equal to $-E_{\mbox{\scriptsize
eff}}=(M_+-M_-)g/2q$.

\subsection{The density profiles} 

In the almost neutral region we can estimate the individual
densities of the particles through the following macroscopic argument.
At altitude $y$ the Coulomb gas pressure $p$ compensates the
gravitational force:
\begin{equation}
\frac{dp}{dy}=-g(\rho_{+}M_++\rho_-M_-)\,.
\end{equation}
This equation, the fact that $\rho_+=\rho_-$, and the equation of
state\citenote{HH} for the two-dimensional two-component plasma for
$\Gamma\leq2$
\begin{equation}
\beta p=(\rho_++\rho_-)\left(1-\frac{\Gamma}{4}\right)\,,
\end{equation}
yield the barometric law
\begin{equation}
\label{baroGen}
\rho_+(y)=
\rho_-(y)
=\rho_0\exp\left[-k_0\left(1-\frac{\Gamma}{4}\right)^{-1}y\right]\,.
\end{equation}
This can be verified in the exact expressions of the densities given by
the Green functions. The dominant term in the Green function $g_{++}$ in
that region is
\begin{equation} 
g_{++}(\r,\r')\sim g_{++}^b(\r,\r')=\frac{m_0}{\pi}\exp[ik_-x/2]\sqrt{uu'}
\int_0^{+\infty}\!I_{\nu}(\moko u_<)K_{\nu}(\moko u_>)\cos \nu
k_0x\,d\nu\,,
\end{equation}
and $g_{--}$ is the same as $g_{++}$ except for the phase factor
$\exp[ik_-x/2]$ that is changed into $\exp[-ik_+x/2]$, which
is in fact irrelevant for the physical quantities.  The integral over
$\nu$ can be approximated by a sum over integers by using
the Euler--MacLaurin expansion when $m_0\gg k_0$.
Then the sum can be performed by using the addition
theorems for the Bessel functions. This gives, for the first two terms of
the expansion,
\begin{eqnarray}
g_{++}^b(\r,\r')\sim&
\lefteqn{\frac{m_0}{\pi}\exp[ik_-x/2]e^{-k_0(y+y')/2}}\qquad
\nonumber\\
\label{gbulk}
&\lefteqn{\times\Bigg[
}&
K_0\left(\moko\left|e^{k_0(ix-y)}-e^{-k_0y'}\right|\right)\\
&&\quad{}+\frac{1}{6}K_0(\moko\exp[-k_0y])K_0(\moko\exp[-k_0y'])
+\cdots\,\,\Bigg]\,.\nonumber
\end{eqnarray}
As stated in section \ref{Model}, when the diameter $R$ of the particles
goes to $0$, the density diverges, so we must cut-off
expression~(\ref{gbulk}) when $|\r-\r'|\to0$ replacing $|\r-\r'|$ by $R$
for $|\r-\r'|\leq R$. Then the regularized form of the particle density
is
\begin{equation}
\label{baro}
\rho_{+}(\r)=\rho_{-}(\r)\sim\frac{m(\r)^2}{2\pi}
K_0(m(\r)R)
\,.
\end{equation}
If $m_0R\ll \exp(-k_0y)$ then
\begin{equation}
\label{realbaro}
\rho_+(\r)=\rho_-(\r)\sim \rho_0e^{-2k_0y}\,,
\end{equation}
where $\rho_0=m_0^2K_0(m_0R)/(2\pi)$ is the density when $\mbf{g}=0$.
Notice that the expression~(\ref{baro}) of $\rho_\pm$ is the same as the
one for $\rho_0$ except that the fugacity~$m(\r)$ depends on the
altitude~$y$.  We have found a barometric law~(\ref{realbaro}) with mass
$M_++M_-$ which agrees with~(\ref{baroGen}) when $\Gamma=2$.

Far away from the neutral zone, for
$y\gg k_0^{-1}\ln(m_0/k_0)$ the dominant term of $\rho_s$ in the case
$m_0\gg k_0$ is
\begin{equation}
\rho_s(y)=\frac{m_0^2 u^2}{\pi^2}
\int_0^{k_{-s}/2k_0}
\sin(\pi\nu)K_\nu(\moko u)^2\,d\nu\,.
\end{equation}
The asymptotic behavior of this integral for $y\to\infty$ is computed in
the Appendix. One finds if $k_s\neq0$
\begin{equation}
\label{densitealinfini}
\rho_s(y)\sim 
\frac{2^{k_{-s}/k_0}}{8\pi^2}
\left[\moko\right]^{k_s/k_0}
k_0\Gamma(k_{-s}/2k_0)^2\sin(\pi k_{-s}/2k_0)\,
\frac{e^{-k_s y}}{y}\,.
\end{equation}
The density $\rho_s$ decays like $y^{-1}\exp(-\beta M_sgy)$. The
argument of the exponential is the same of the one of an ideal neutral
gas, but the decay of the density is faster than the one of an ideal
gas (because of the factor $y^{-1}$).

In real colloidal suspensions the mass of a counterion is much smaller
than the mass of a colloidal particle.  If we assume in our model that
the negative couterions have mass $M_-=0$ then the
expression~(\ref{gss}) for the positive colloidal particles is
simplified: the last integral in~(\ref{gss}) vanishes. The dominant term
of the profile density of the colloidal particles given by~(\ref{baro})
is now valid not only in neutral zone but also up to $y\to\infty$.  For
the counterions the profile density in the neutral zone is still
dominated by~(\ref{baro}) but for $y\gg k_0^{-1}\ln(m_0/k_0)$
expression~(\ref{densitealinfini}) is no longer valid because $k_-=0$.
Instead we have (see the Appendix for details)
\begin{equation}
\label{ocp}
\rho_-(y)\sim \frac{1}{4\pi y^2}\,.
\end{equation}
We recover an already known result: in this particular case where
$k_-=0$, in the considered region, because of the exponential decay of
the density $\rho_+(y)$, one can consider that the system is composed
only of particles with charge $-q$ without mass: the system is a
one-component plasma without a background. The asymptotic behavior
given by~(\ref{ocp}) was found for the solvable model of the
one-component plasma without a background\citenote{Janco}. 
The following macroscopic argument also leads to expression~(\ref{ocp}):
the one-component plasma without a background can be seen as a
neutralized one-component plasma (where we have added a background charge
density $q\rho_-$) immersed in an external charged background $-q\rho_-$.
At altitude $y$ the gradient of the thermal pressure of the neutralized
plasma compensates the electric force exerced on the external
background: 
\begin{equation}
\frac{dp(y)}{dy}=-q\rho_-(y)E(y)\,,
\end{equation}
where $p(y)$ is the thermal pressure of the neutralized one-component
plasma and $E(y)$ is the electric field created by the charge
distribution $-q\rho_-(y)$. The Poisson equation is
\begin{equation}
\frac{dE(y)}{dy}=-2\pi q\rho_-(y)\,.
\end{equation}
These two equations together with the equation of state for the neutralized
one-component plasma
\begin{equation}
\beta p=\left(1-\frac{\Gamma}{4}\right)\rho_-\,,
\end{equation}
yield the differential equation for $\rho_-$
\begin{equation}
\left(1-\frac{\Gamma}{4}\right)\frac{d}{dy}\left[
\frac{1}{\rho_-(y)}\frac{d\rho_-(y)}{dy}\right]
=2\pi\Gamma\rho_-(y)\,,
\end{equation}
with the solution
\begin{equation}
\rho_-(y)=\frac{1-\Gamma/4}{\Gamma\pi(y-y_0)^2}%
\mathop{\sim}\limits_{y\to\infty}
\frac{1-\Gamma/4}{\Gamma\pi y^2}\,.
\end{equation}
which agrees with~(\ref{ocp}) for $\Gamma=2$.

Let us now consider that the mass of the counterions is not zero but
that $k_-\ll k_+$. In that case equation~(\ref{densitealinfini}) becomes
for the counterions
\begin{equation}
\label{ocp-E}
\rho_-(y)\sim 
\frac{k_-}{4\pi}
\frac{e^{-k_s y}}{y}\,.
\end{equation}
It is interesting to notice that this expression is the same 
as the one which would be
obtained
for a one-component plasma without a background but with massive particles.
For a one-component plasma, it is
equivalent to consider that the particles have a mass $M_-$ or that they
are in an external electric field $E=-M_-g/q$ and do not have a
mass.  The calculations done in Ref.~\cite{Janco} can be easily extended
to the present case and they give the same
asympotic behavior~(\ref{ocp-E}) for the particle density.

\section{Conclusion}

We have discussed the density profiles in this model of sedimentation
equilibrium. Unfortunately our model cannot account for excluded volume
interactions which are also important in the sedimentation equilibrium.
Nevertheless,
for heights of the container of the same order as the
gravitational length, we find the same qualitative results as in other
models\citenote{BibHans}: heavier particles accumulate at the bottom
while lighter particles accumulate at the top. For heights of the
container much larger than the gravitational length, we find an
interesting phenomenon: lighter particles do not accumulate any longer
at the top of the container but there is an excess of them forming a
cloud centered at an
altitude of order of the gravitational length times some logarithm. An
open problem is whether such a cloud of light charged particles might be
observed, because we have not considered possible instabilities in the
horizontal direction, with the excess charge going to the lateral walls
for instance.

\section*{Acknowledgments}

I wish to thank B.~Jancovici for useful discussions and for critical
reading of the manuscript.

\setcounter{equation}{0}
\renewcommand{\theequation}{A.\arabic{equation}}
\section*{Appendix}

We wish to find the asymptotic behavior of the integral
$\int_0^{k_s/2k_0} \sin(\pi\nu)K_\nu(m_0 u/k_0)^2\,d\nu$ when
$u\to0$ ($y\to\infty$). We define $z=m_0 u/k_0$, $\alpha=k_s/2k_0$,
and
\begin{equation}
\label{defIz}
I(z)=
\int_0^\alpha \sin(\pi\nu) K_\nu(z)^2\,d\nu\,.
\end{equation}
We use the formula
\begin{equation}
K_\nu(z)^2=2\int_0^\infty K_0(2z\cosh t) \cosh(2\nu t)\,dt \,,
\end{equation}
in equation~(\ref{defIz}) and perform the integral over $\nu$ to find
\begin{equation}
\label{Iz}
I(z)=
2
\int_0^\infty
\frac{%
\pi-\pi\cos(\alpha\pi)\cosh(2\alpha t)
+2t\sin(\alpha\pi)\cosh(2\alpha t)}%
{\pi^2+4t^2}
K_0(2z\cosh t)\,dt\,.
\end{equation}
If $\alpha<1$ (ie. if $k_{-s}\neq0$) then doing the change of variable
$s=2z\cosh t$ in~(\ref{Iz}) one finds that
\begin{equation}
\lim_{z\to0} z^{2\alpha}\ln(1/z)I(z)=
2^{-1}\sin(\alpha\pi)
\int_0^\infty s^{2\alpha-1}K_0(s)\,ds\,.
\end{equation}
The integral over $s$ can be done and finally we have
\begin{equation}
I(z)\mathop{\sim}\limits_{z\to0}
\frac{4^{\alpha}}{8}\Gamma(\alpha)^2\sin(\alpha\pi)
\,\frac{z^{-2\alpha}}{\ln(1/z)}\,.
\end{equation}
This expression is not valid when $\alpha=1$, that is in the case when one
of the masses of the particles is zero. In that case equation~(\ref{Iz})
becomes 
\begin{equation}
I(z)=
4
\int_0^\infty
\frac{\pi}{\pi^2+4t^2}
K_0(2z\cosh t)\cosh^2 t\,dt\,.
\end{equation}
Then doing again the change of variable $s=2z\cosh t$ one finds
\begin{equation}
I(z)\mathop{\sim}\limits_{z\to0}
\frac{\pi}{4}\,\frac{1}{z^2\ln(1/z)^2}\,.
\end{equation}

\newpage

\end{document}